\documentclass[aps,prd,preprint,preprintnumbers,unsortedaddress,superscriptaddress,showpacs,nofootinbib]{revtex4-1}

\pdfoutput=1
\usepackage{graphicx}
\usepackage{amsmath}
\usepackage{amsfonts}
\usepackage{amssymb}
\usepackage{color}

\usepackage[colorlinks, citecolor=blue,anchorcolor=red,menucolor=red, linkcolor=red,filecolor=red,runcolor=red,urlcolor=blue,frenchlinks=red]{hyperref}

\usepackage{bm}
\usepackage[section]{placeins}
\usepackage{booktabs}

\begin{document}
\arraycolsep1.5pt

\title{Polarization amplitudes in $\tau^- \to \nu_{\tau} V P$  decay \\beyond the Standard Model}


\author{L.~R.~Dai}
\email{dailr@lnnu.edu.cn}
\affiliation{Department of Physics, Liaoning Normal University, Dalian 116029, China}
\affiliation{Departamento de F\'isica Te\'orica and IFIC, Centro Mixto Universidad de Valencia-CSIC,
Institutos de Investigac\'ion de Paterna, Aptdo. 22085, 46071 Valencia, Spain
}

\author{E. Oset}
\email{oset@ific.uv.es}
\affiliation{Departamento de F\'isica Te\'orica and IFIC, Centro Mixto Universidad de Valencia-CSIC,
Institutos de Investigac\'ion de Paterna, Aptdo. 22085, 46071 Valencia, Spain
}

\date{\today}
\begin{abstract}
We study the amplitudes  of $\tau^- \to \nu_{\tau} V P$  decay for the different polarizations of the vector meson $V$, using a  formalism   where the mapping  from the
quark degrees of freedom  to the  meson ones is done  with the $^3{P}_0$ model.
 We extend the  formalism  to a  case, with the operator $\gamma^\mu -\alpha\gamma^\mu \gamma_5$, that can account for different models beyond the Standard Model£¬
 and study in detail the  $\tau^- \to \nu_{\tau} K^{*0} K^{-}  $ reaction for the different  polarizations of the $K^{*0}$. The results are shown in
 terms of the $\alpha$ parameter that differs for each model. We find that  $\frac{d \Gamma}{d M_{\rm inv}^{(K^{*0} K^{-} )}}$  is very different for
 each of third components of the vector spin, $M=\pm 1, 0$,  and in particular the magnitude
  $\frac{d \Gamma}{d M_{\rm inv}^{(K^{*0} K^{-} )}}|_{M=+1} -\frac{d \Gamma}{d M_{\rm inv}^{(K^{*0} K^{-} )}}|_{M=-1} $ is very sensitive to the $\alpha$ parameter, which
  makes the investigation of this  magnitude very useful  to test different models beyond the standard model.
 \end{abstract}

\maketitle

\date{\today}

\maketitle
\section{Introduction}
The $\tau$ decays have received much attention through the years and have proved to be a good source of information on weak interaction, and also strong interaction from the hadronic
decay modes \cite{isgur89,npb84,PRe88,Rep2006,PRe2010,bali,bali2}.  While most of the attention in  hadronic decays is given to the modes $\tau^- \to \nu_\tau M$, $\tau^- \to \nu_\tau M_1 M_2$,
with $M_1 M_2$ pseudoscalar mesons,  much less is known about $\tau^- \to \nu_\tau V M $ decays, with $V$ a vector and $M$  a pseudoscalar, although  several modes are measured, $\tau^- \to \nu_{\tau} \omega \pi^{-}  $ \cite{Buskulic1,Buskulic2}, $\tau^- \to \nu_{\tau} K^{*0} K^{-}$ \cite{Barate},   $\tau^- \to \nu_{\tau} K^{*-}  \eta $ \cite{Inami},  $\tau^- \to \nu_{\tau} \omega K^{-}  $ \cite{Arms},
$\tau^- \to \nu_{\tau}  \rho^{-} \pi^{0}$ \cite{Barate2}, $\tau^- \to \nu_{\tau} K^{*0} \pi^{-}  $ \cite{Barate},
$\tau^- \to \nu_{\tau} \phi \pi^{-} $  \cite{Aubert}.  Theoretically there have been also a few works dealing with these reactions. The $\tau^- \to \nu_{\tau} \rho \eta $
decay is studied in \cite{volroeta} using the extended the Nambu$-$Jona-Lasinio  model.  The  $\tau^- \to \nu_{\tau} K^{*} \bar{K} $ decay is evaluated in \cite{bali} using vector
meson dominance. The  $\tau^- \to \nu_{\tau} \omega \pi^{-}  $  decay is also addressed in \cite{bali,bali2} and in \cite{castro,volpiom}.
In Ref. \cite{bali,bali2} the $\tau^- \to \nu_{\tau} (\rho^- \bar{K}^0  + \rho^0 K^-)$ mode is also investigated, together with the $\tau^- \to \nu_{\tau}  \omega K^{-}$  and
$\tau^- \to \nu_{\tau}  K^{*} \eta$ modes.  The widths are obtained summing over the polarizations of the vector meson.

A different approach to the  $\tau^- \to \nu_{\tau} M_1 M_2$, $\tau^- \to \nu_{\tau} V M$, $\tau^- \to \nu_{\tau} V V$ reactions is done in \cite{daitau} where the $\tau^- \to \nu_{\tau}  \bar{u} d$ ($\bar{u} s$
for Cabibbo-suppressed decay)  primary process is considered and the $\bar{u} d$ pair is hadronized  inserting an extra $\bar{q}q$  pair using the $^3{P}_0$ model \cite{micu,oliver,close}.
The novelty of the approach of  \cite{daitau}  is that an elaborate angular momentum algebra calculation is performed that allows one to relate the different  processes, up to a global
form factor from the matrix element of the quark radial wave functions. This approach  is in line with the one followed for the $B \to M_1 M_2$ weak decay in \cite{liang18}. As far as this
form factor  is similar for  $\tau^- \to \nu_{\tau} K^{*0} \bar{K} $ or $\tau^- \to \nu_{\tau} K^{*0} \bar{K}^* $ the ratio of the decay rates can be calculated, and  one can also relate  cases like
$\tau^- \to \nu_{\tau} K^{*0} K^{-}$ with $\tau^- \to \nu_{\tau} \rho^{0}  K^{-} $, $\tau^- \to \nu_{\tau} \omega K^{-}$, $\tau^- \to \nu_{\tau} K^{*-} \eta $ etc. Except in cases with $\pi$ production where the
form factors are rather different than for other processes, the results obtained are in fair agreement with experiments.

Another good feature of this approach is that it provides directly the contribution for each value of the vector spin polarization. In \cite{daitau}  and  \cite{bali,volroeta,castro,volpiom} the sum over
polarizations of $\tau,\nu_\tau$ and the vector is performed. The purpose of the present work is to evaluate explicitly the contribution of each polarizations component, and
we shall see that they are rather different, and in view of this, we perform  the calculations for the  weak operator of the hadronic vertex $\gamma^\mu -\alpha\gamma^\mu \gamma_5$
[$\alpha=1$ for the Standard Model (SM)] that one encounters in models beyond the  Standard Model (BSM), and conclude that
the experimental investigation  of these magnitudes  can bring valuable information concerning  extrapolations of the Standard Model. Although
this type of Lagrangian does not include the different extrapolations BSM \cite{gang,avelino,okada,new1,darice,bordone,Fornal,new2,German}. (see also the mini-review ``Muon decay parameters" in PDG \cite{pdg}),
the results obtained for this  sub-set of BSM models is rather illustrative, as we shall see.

 Interestingly, such studies have been conducted  in related reactions, as the semileptonic decay of mesons, like the $B \to  D^{*} \bar{\nu} l$. Indeed,  helicity amplitudes
 are evaluated  in \cite{wade} and longitudinal  and transverse polarizations are also separated in the  $B \to  D^{*} \bar{\nu}_\tau \tau$ decay in \cite{jorge}. In \cite{helicity}
we performed also the calculations of the $D^*$ helicity amplitudes for the $B \to  D^{*} \bar{\nu} l$ decay and extrapolated  them for the general case of $\gamma^\mu -\alpha\gamma^\mu \gamma_5$
weak quark vertex, varying the value of $\alpha$  to accommodate  potential models BSM. We found that the results, in particular  the difference of contributions from $M=-1$ and $M=+1$, were very sensitive
to $\alpha$, rendering  this magnitude  a very good instrument to advance in our comprehension of the Standard Model and beyond.

The present work follows the idea of \cite{helicity} and also evaluates the contributions of $M=0,\pm 1$, the third component of the $K^{*0}$ spin along the direction of the  $\nu_{\tau} $,
for the $\tau^- \to \nu_{\tau} K^{*0} K^{-}$  reactions. The approach of  \cite{daitau}  is most suited  to this study since what is missing  in this approach is a form
factor coming from the integral of $j_0 (qr)$  with the radial part of the linear quark wave functions involved,  with $q$ the momentum transfer. Yet, this form
factor cancels exactly in the ratios of amplitudes, which in our approach are then produced without any free parameter.  As in \cite{helicity} here we also find a big
sensitivity of the ratios of the polarization amplitudes to the value of $\alpha$  in  $\gamma^\mu -\alpha\gamma^\mu \gamma_5$, which converts the study of these magnitudes into an excellent tool to further learn about
the weak interaction and the possible influence of new physics.

\section{Formalism}

\begin{figure}
 \begin{center}
	\includegraphics[width=0.5\textwidth]{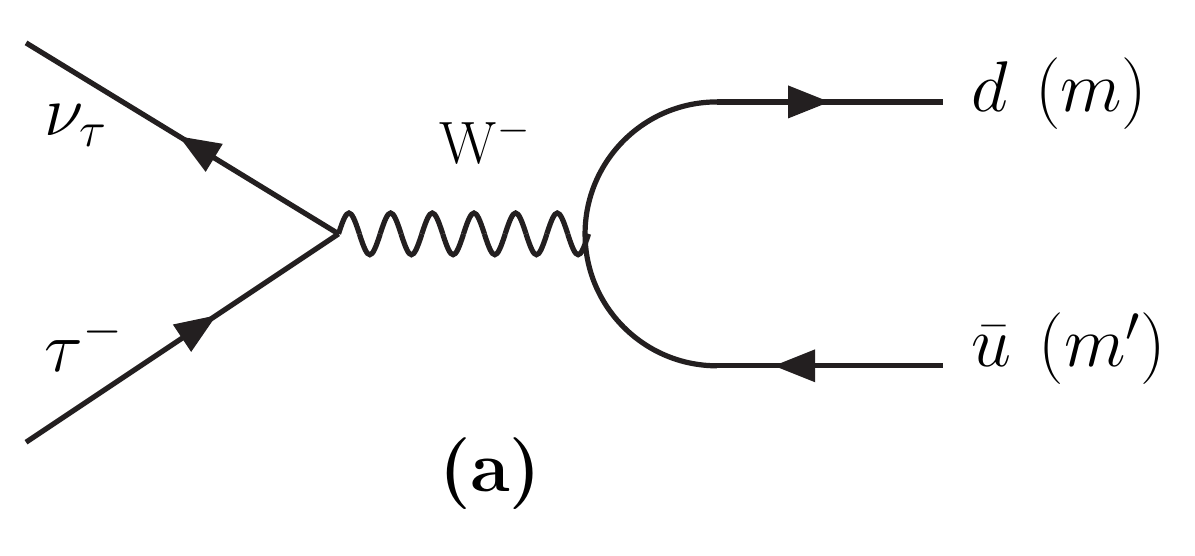}	\includegraphics[width=0.54\textwidth]{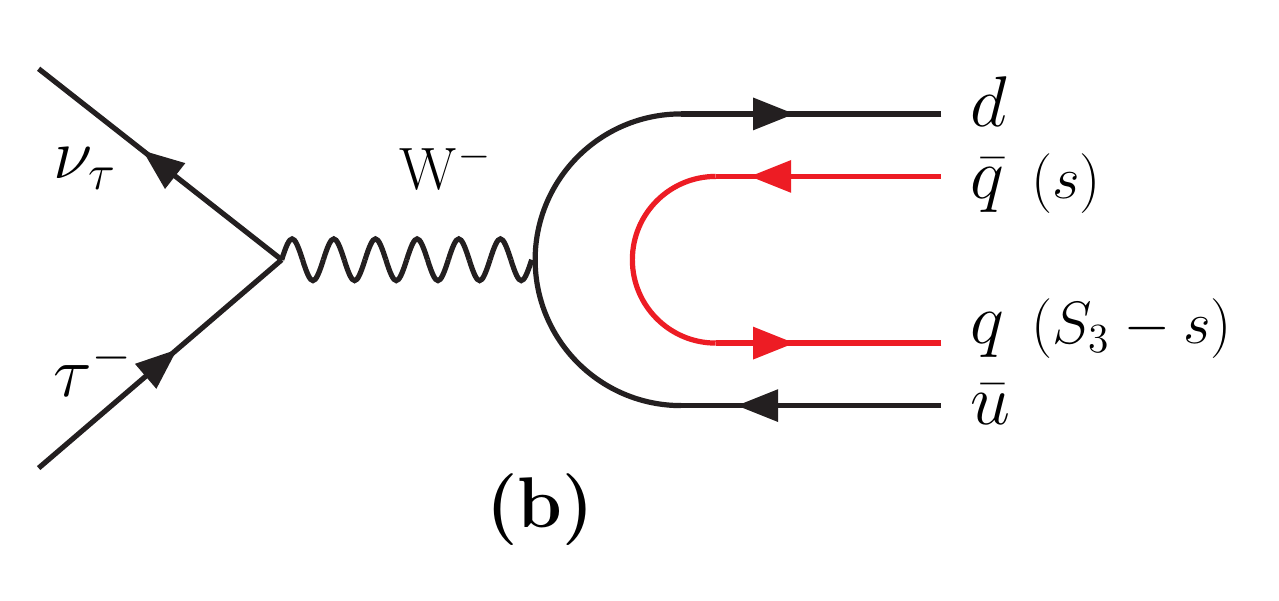}	
	\end{center}
	\caption{\label{fig:tau} (a) Elementary $\tau^- \to \nu_{\tau} d \bar{u}$ diagram. The labels $m,m^{\prime}$ stand for the third component of spin of the quarks;
(b) Hadronization  of the primary $d\bar{u}$ pair to produce two mesons, $s$ is the third component of the spin of $\bar{q}$ propagating as a particle, while
$S_3-s$ is the third component of the spin of $q$ , where  $S_3$ is the third component of the total  spin $S$ of $\bar{q}q$. }
\end{figure}

We will study the polarization amplitudes  for $\tau^- \to \nu_{\tau} K^{*0} K^{-}$. As shown in  \cite{daitau}, the reaction proceeds in $s$-wave for $K^{*0} K^{-}$ and the results correspond
to the case  $J=1,J'=0$ studied there with $J$ the spin of the $K^{*0}$  and $J'$ the one of $K^{-}$. Diagrammatically the reaction proceeds as shown in Fig. \ref{fig:tau}. The $W$ produces a
$d \bar{u}$ quark pair ($s\bar{u}$ in the Cabibbo suppressed case) which   hadronizes due to  the creation of a $\bar{q}q$ pair with the quantum numbers of the vacuum.
The hadronization is taken into account by means of the $^3{P}_0$ model, and by taking $\bar{q}q=\bar{u}u+ \bar{d}d+\bar{s}s$  we can relate different flavors in the  $\tau^- \to \nu_\tau V M$ \cite{daitau}.
However, since we only wish to investigate  $\tau^- \to \nu_{\tau} K^{*0} K^{-}$, it is sufficient to see that this mode is obtained   hadronizing $d\bar{u}$ with $\bar{s}s$.
The elementary quark interaction  is given by
\begin{eqnarray}
H= \mathcal{C} L^\mu Q_\mu \,,
\end{eqnarray}
where the $\mathcal{C}$ contains the couplings of the weak  interaction. The constant $\mathcal{C}$  plays no role in our study because we are only concerned about ratios of rates.
 The  leptonic current is given by
\begin{eqnarray}
L^\mu=\langle {\bar u}_\nu |\gamma^\mu-\gamma^\mu\gamma_5| u_\tau\rangle \,,
\end{eqnarray}
and  the quark current  by
\begin{eqnarray}
Q^\mu=\langle \bar u_d|\gamma^\mu-\gamma^\mu\gamma_5|v_{\bar u}\rangle  \,.
\end{eqnarray}
As is usual in the evaluation of decay widths to three final particles, we evaluate the matrix elements in the frame where the two mesons system is at rest.
For the evaluation of the matrix element $Q_\mu$ we assume that the quarks are at rest in that frame and we have
in the Itzykson-Zuber normalization \cite{itzy} the  spinors
\begin{equation}\label{eq:wfn}
u_r=
\left(
\begin{array}{c}
\chi_r\\ 0
\end{array}
\right),
v_r=
\left(
\begin{array}{c}
0\\ \chi_r
\end{array}
\right),
\chi_1=
\left(
\begin{array}{c}
1\\ 0
\end{array}
\right),
\chi_2=
\left(
\begin{array}{c}
0\\ 1
\end{array}
\right),
\end{equation}
\begin{eqnarray} \label{eq:Qu2}
Q_0&=& \langle \chi^{\prime} | 1 | \chi \rangle \equiv M_0  \nonumber \, , \\
Q_i&=&  \langle \chi^{\prime} | \sigma_i |\chi  \rangle \equiv N_i \, .
\end{eqnarray}

Even if the mesons are at rest, the quarks will have some internal momentum. This would require  to use the  small components in the spinors, but they contribute as $(p_{in}/2m)^2$, and
with values of  $p\simeq 200~ {\rm MeV}/c$ and  an average constituent mass of $400$ MeV the effects are of the order of $6\%$, which are assumed in  this approach.  Small as these effects are,
they  further  tend to cancel  in ratios  of amplitudes, such that  we should not worry  about them.

To obtain the $\tau$ width, we  must evaluate
\begin{eqnarray}\label{eq:L}
\overline{\sum} \sum  L^\mu {L^\nu}^\dagger Q_\mu Q_\nu^*= \bar{L}^{00}\, M_0~ M^*_0
+\bar{L}^{0i}\,M_0 ~N^*_i
+\bar{L}^{i0} \, N_i ~M^*_0
+\bar{L}^{ij} N_i ~N_j^* \, .
\end{eqnarray}
Denoting for simplicity,
\begin{eqnarray}\label{eq:LL}
\bar{L}^{\mu\nu}&=& \overline{\sum} \sum  L^\mu {L^\nu}^\dagger  \nonumber \, \\
&=& \frac{1}{m_{\nu} m_{\tau}}\left( p'^\mu p^\nu  +p'^\nu p^\mu - g^{\mu\nu}p'\cdot p+i \epsilon^{\alpha\mu\beta\nu}p'_\alpha p_\beta\right) \, ,
\end{eqnarray}
where $p,p'$ are the momenta of the $\tau$ and $\nu_\tau$ respectively and  we use the field normalization for fermions of Ref. \cite{mandl}.

For the  $J=1, J'=0$ (vector- pseudoscalar) case, we obtain  the results \cite{daitau}:
\begin{equation}\label{eq:m0}
M_0=\frac{1}{\sqrt{6}}\frac{1}{4\pi} \, \delta_{M'0}
\end{equation}
\begin{equation}\label{eq:nu}
N_\mu= (-1)^{-\mu} \frac{1}{\sqrt{3}}\frac{1}{4\pi} {\cal C}(1 1 1; M,-\mu,M-\mu)\,\delta_{M'0}
\end{equation}
with ${\cal C}(\cdots)$ a Clebsch-Gordan coefficient. The $\mu$ index of Eq. \eqref{eq:nu} is  the index of $N_i$ in spherical basis
\begin{equation}\label{eq:spbs}
N_{+1}=-\frac{1}{\sqrt{2}}(N_1+iN_2), ~~~ N_{-1}=\frac{1}{\sqrt{2}}(N_1 - iN_2), ~~~N_0=N_3  \, .
\end{equation}
This formalism  is demanded when one has to project over spin components. In addition, $M, M'$ are the third components of the $K^{*0}$ and $K^-$  respectively (obviously $M'=0$, do not
confuse $M, M'$ with the amplitude $M_0$ of Eqs.\eqref{eq:Qu2},\eqref{eq:L} and \eqref{eq:m0}).
The quantization axis is taken along the direction of the neutrino in the $\tau^-$ rest frame.

Using Eqs. \eqref{eq:L},\eqref{eq:LL} and  \eqref{eq:m0},\eqref{eq:nu} and following the steps of the appendix A,  we obtain  the $\tau$ decay amplitude, $t$,
 up to a global constant, and we find
\begin{itemize}
\item[1)] $M=0$
\begin{eqnarray}\label{eq:tM0}
\overline{\sum} \sum \left|t\right|^2= \frac{1}{m_\tau m_\nu} \frac{1}{6}\left(\frac{1}{4 \pi}\right)^2 \,\left(3 E_\tau E_\nu - p^2 \right) \, ,
\end{eqnarray}
\item[2)] $M=1$
\begin{eqnarray}\label{eq:tM1}
\overline{\sum} \sum \left|t\right|^2= \frac{1}{m_\tau m_\nu} \frac{1}{6}\left(\frac{1}{4 \pi}\right)^2 \,\left[3 E_\tau E_\nu + p^2 +(3 E_\nu +E_\tau) p  \right] \, ,
  \end{eqnarray}
 \item[3)] $M=-1$
\begin{eqnarray}\label{eq:tMm1}
\overline{\sum} \sum \left|t\right|^2= \frac{1}{m_\tau m_\nu} \frac{1}{6}\left(\frac{1}{4 \pi}\right)^2 \,\left[3 E_\tau E_\nu + p^2 -(3 E_\nu +E_\tau) p  \right] \, ,
 \end{eqnarray}
 \end{itemize}
where $p$ is the momentum of the $\tau$, or $\nu_\tau$, in the $K^{*0}K^- $ rest frame, given by
\begin{equation}\label{eq:newlabel}
p=p_\nu=p_\tau=\frac{\lambda^{1/2}(m^2_\tau,m^2_\nu,M_{\rm inv}^{2 (K^{*0} K^-)})}{2 M_{\rm inv}^{(K^{*0} K^-)}}\, ,
\end{equation}
$E_\tau=\sqrt{m_\tau^2+p^2}$, $E_\nu=p$ and $\bar{L}^{\mu\nu}$ of Eq. \eqref{eq:LL} is evaluated in this frame too.


\section{Results}

In the former equations the angle integrations are already  done in a way that finally we must take into account the full phase space with the angle independent expressions. So, we have
\begin{equation}\label{eq:dGdM}
\frac{ d\Gamma}{d M_{\rm inv}^{(K^{*0} K^-)}} =  \frac{2\,m_\tau 2\, m_\nu}{(2\pi)^3} \frac{1}{4 m^2_\tau}\, p_\nu {\widetilde p_1}\, \overline{\sum} \sum \left|t\right|^2 \,,
\end{equation}
where $p_\nu$ is the neutrino momentum in the $\tau$ rest frame
\begin{equation}
p_\nu=\frac{\lambda^{1/2}(m^2_\tau,m^2_\nu,M_{\rm inv}^{2 (K^{*0} K^-)})}{2 m_\tau}\, ,
\end{equation}
and  $ {\widetilde p_1}$ the momentum of $K^{*0}$  in the $K^{*0}K^-$ rest frame  given by
\begin{equation} \label{eq:new2}
\widetilde{p}_1=\frac{\lambda^{1/2}(M_{\rm inv}^{2 (K^{*0} K^-)}, m_{K^{*0}}^2, m_{K^-}^2)}{2 M_{\rm inv}^{(K^{*0} K^-)}}\, .
\end{equation}
The total  differential width is given by
\begin{eqnarray}\label{eq:R}
 R=\frac{d \Gamma}{d M_{\rm inv}^{(K^{*0} K^-)}}|_{M=+1} + \frac{d \Gamma}{d M_{\rm inv}^{(K^{*0} K^-)}}|_{M=0} +\frac{d \Gamma}{d M_{\rm inv}^{(K^{*0} K^-)}}|_{M=-1} \, .
\end{eqnarray}

\begin{figure}[ht]
\includegraphics[scale=0.8]{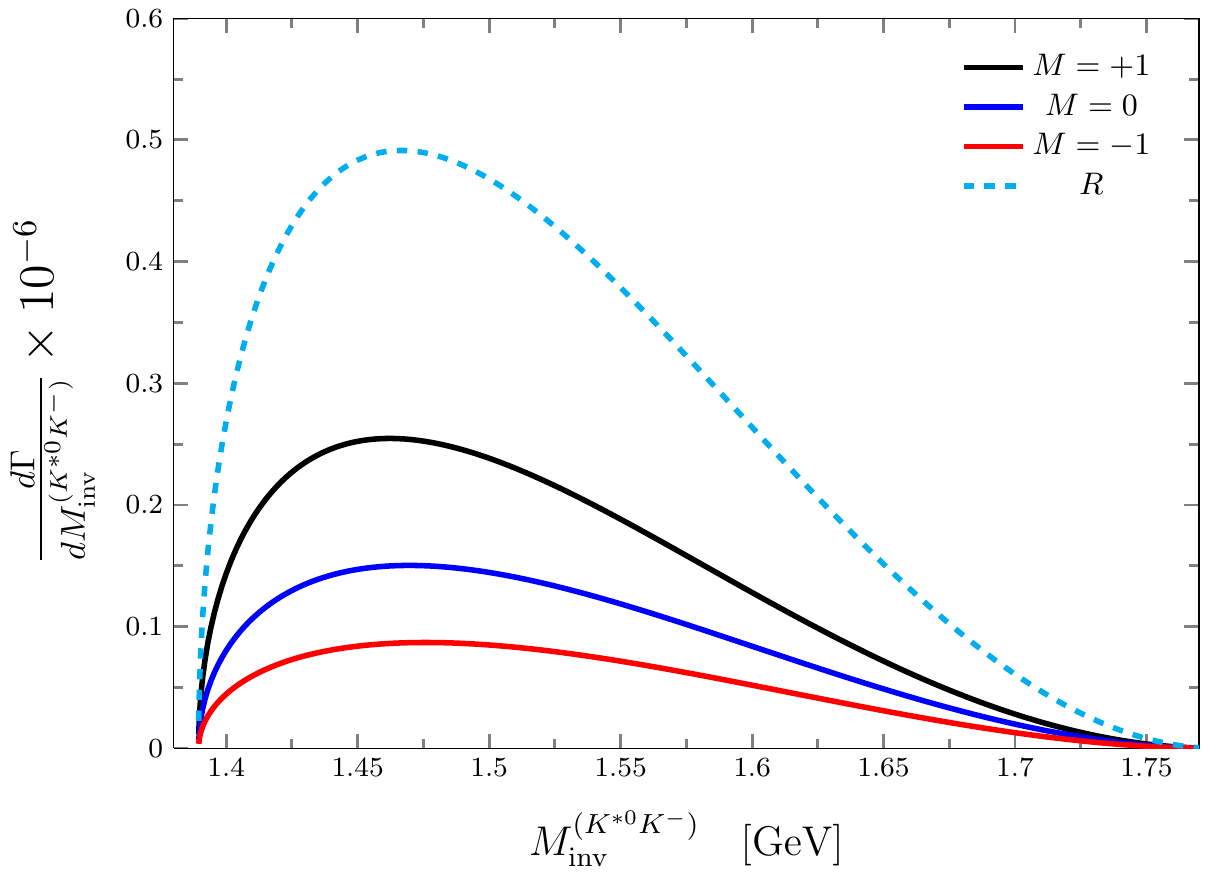}
\caption{Total  differential width $R$ of Eq. \eqref{eq:R}, and individual contributions  of $\frac{d \Gamma}{d M_{\rm inv}^{(K^{-} K^{*0})}}|_{M=0}$,
$\frac{d \Gamma}{d M_{\rm inv}^{(K^{-} K^{*0})}}|_{M=-1}$, and $\frac{d \Gamma}{d M_{\rm inv}^{(K^{-} K^{*0})}}|_{M=+1}$.
}
\label{fig:dg}
\end{figure}

\begin{figure}[ht]
\includegraphics[scale=0.8]{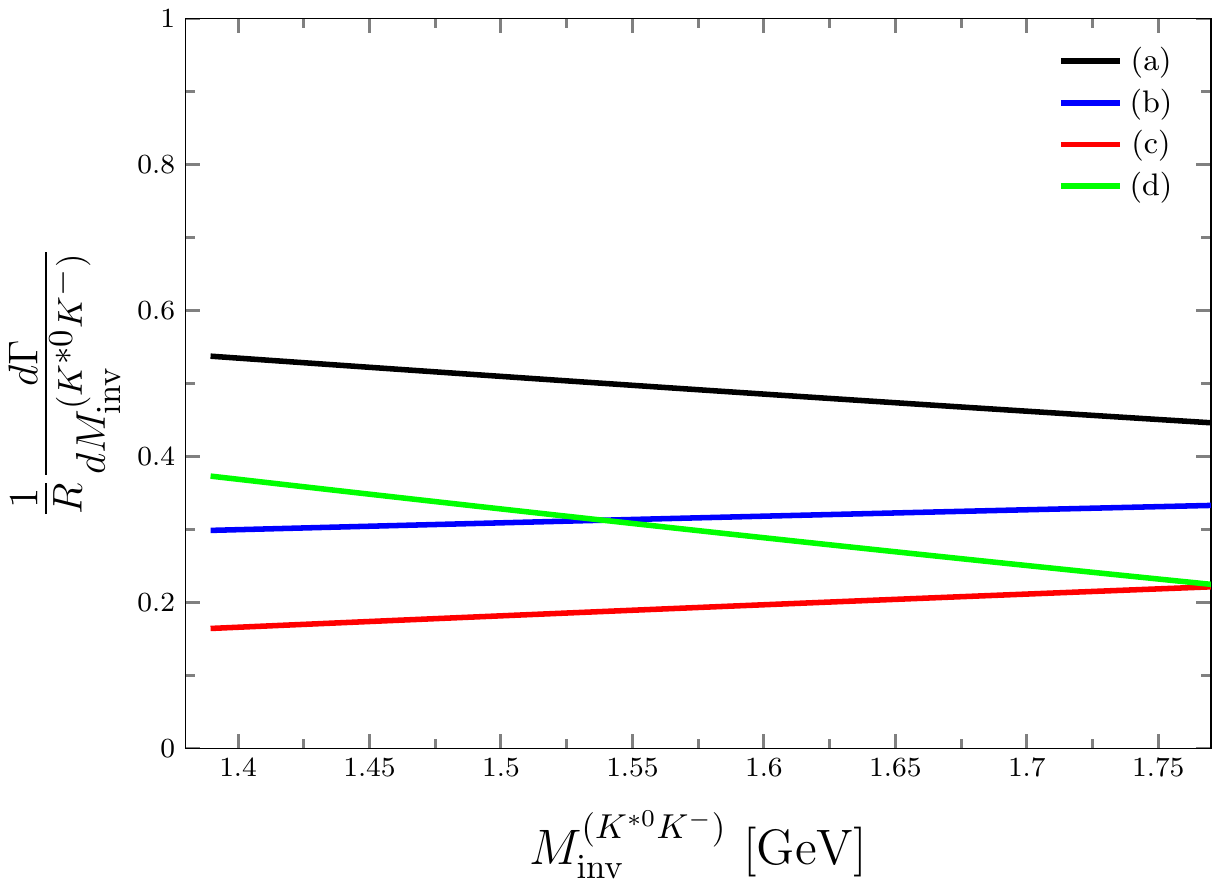}
\caption{The different  ratios, where lines (a), (b) and (c) show  $\frac{d \Gamma}{d M_{\rm inv}^{(K^{-} K^{*0})}}|_{M=+1}$, $\frac{d \Gamma}{d M_{\rm inv}^{(K^{-} K^{*0})}}|_{M=0}$, and
$\frac{d \Gamma}{d M_{\rm inv}^{(K^{-} K^{*0})}}|_{M=-1}$ respectively, and line (d) denotes the
 difference of  $\frac{d \Gamma}{d M_{\rm inv}^{(K^{-} K^{*0})}}|_{M=+1}-\frac{d \Gamma}{d M_{\rm inv}^{(K^{-} K^{*0})}}|_{M=-1}$, all  divided by the total differential width $R$ of Eq. \eqref{eq:R}.
 }
\label{fig:ro}
\end{figure}

In Fig. \ref{fig:dg} we  show the individual contribution of each $M$ and the total $R$. In Fig. \ref{fig:ro} we show the contribution of each $M$ and the difference
of $M=+1$  and $M=-1$, divided by  the total  differential width  $R$.

In the search for contributions BSM one  usually compares some magnitude with experiment and diversions of experiment with respect to the SM predictions
 are seen as a signal of possible new physics. So far the experimental errors do not make  the cases compelling. The present case could offer a good opportunity,  since
the individual contributions for different $M$ vary appreciably when diverting from the Standard Model, as we show in the next section.



\section{Consideration of right-handed quark currents}
There is a huge amount of work on extensions of the Standard Model and this is not the place to discuss it. We only like to mention current models which are widely used recently,
as minimal gauge extensions of the SM \cite{gang,avelino}, leptoquarks\cite{okada}, scalar leptoquarks \cite{new1,darice},
Pati-Salam gauge models \cite{bordone,Fornal,new2} and right-handed models \cite{German,hedos}.

Some models BSM have quark currents that contain the combination $\gamma^\mu+ \gamma^\mu \gamma_5$.
Some of  the models mentioned above could be accounted for with an operator
\begin{eqnarray}
&a(\gamma^\mu -\gamma^\mu \gamma_5)+b(\gamma^\mu+\gamma^\mu \gamma_5) \nonumber \, \\
=&(a+b)\left\{\gamma^\mu -\frac{a-b}{a+b}\gamma^\mu \gamma_5 \right\}  \nonumber \,.~~~~
\end{eqnarray}
We shall call $\frac{a-b}{a+b}=\alpha$ and  study the   distributions  for different $M$ as a function of $\alpha$.  Thus, we  have the operator
\begin{eqnarray}\label{eq:19}
\gamma^\mu -\alpha\gamma^\mu \gamma_5   \,  .
\end{eqnarray}

\begin{figure}[ht!]
\includegraphics[scale=0.82]{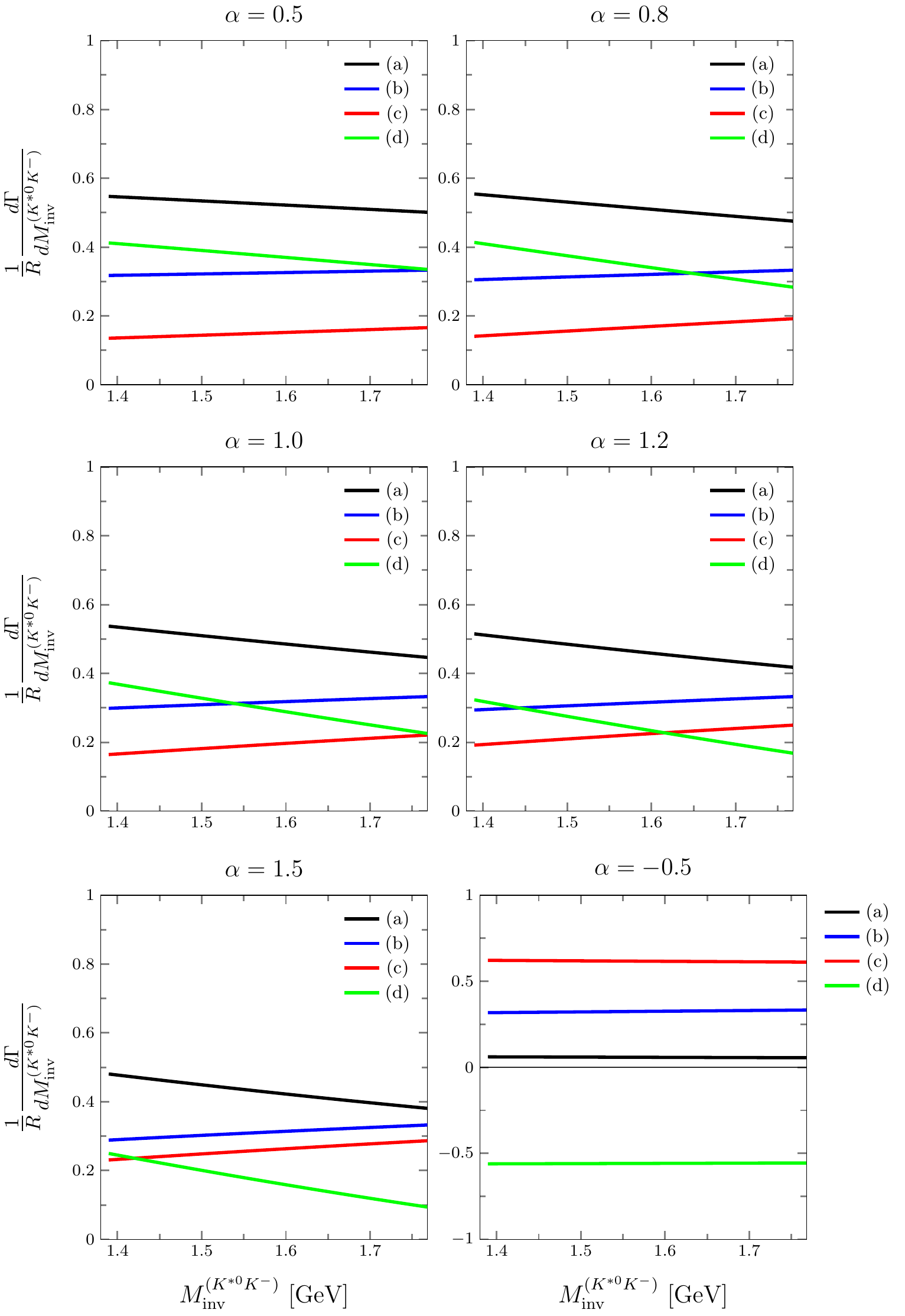}
\caption{The same as Fig. \ref{fig:ro} but for different  $\alpha$.  }
\label{fig:alpha}
\end{figure}

Using the same formalism  of \cite{daitau}, and the appendix A of the present work,  it is easy to see the results as a function of $\alpha$. We obtain the following results:
\begin{itemize}
\item[1)] $M=0$
\begin{eqnarray}\label{eq:tM0}
\overline{\sum} \sum \left|t\right|^2= \frac{1}{m_\tau m_\nu} \frac{1}{6}\left(\frac{1}{4 \pi}\right)^2 \,\left\{ \left(E_\tau E_\nu + p^2 \right) + 2 \alpha^2 \left(E_\tau E_\nu - p^2 \right) \right\}\, ,
\end{eqnarray}
\item[2)] $M=1$
\begin{eqnarray}\label{eq:tM1}
\overline{\sum} \sum \left|t\right|^2 &=& \frac{1}{m_\tau m_\nu} \frac{1}{6}\left(\frac{1}{4 \pi}\right)^2 \,\left\{(E_\tau E_\nu + p^2) + 2 \alpha (E_\nu +E_\tau) p  \right.\nonumber \, \\
&+& \left. \left[2E_\tau E_\nu +(E_\nu -E_\tau) p \right] \alpha^2 \right\}
  \end{eqnarray}
 \item[3)] $M=-1$
\begin{eqnarray}\label{eq:tMm1}
\overline{\sum} \sum \left|t\right|^2 &=& \frac{1}{m_\tau m_\nu} \frac{1}{6}\left(\frac{1}{4 \pi}\right)^2 \,\left\{(E_\tau E_\nu + p^2) - 2 \alpha (E_\nu +E_\tau) p  \right.\nonumber \, \\
&+& \left. \left[2E_\tau E_\nu -(E_\nu -E_\tau) p \right] \alpha^2 \right\}
 \end{eqnarray}
 \end{itemize}

\begin{figure}[ht!]
\includegraphics[scale=0.7]{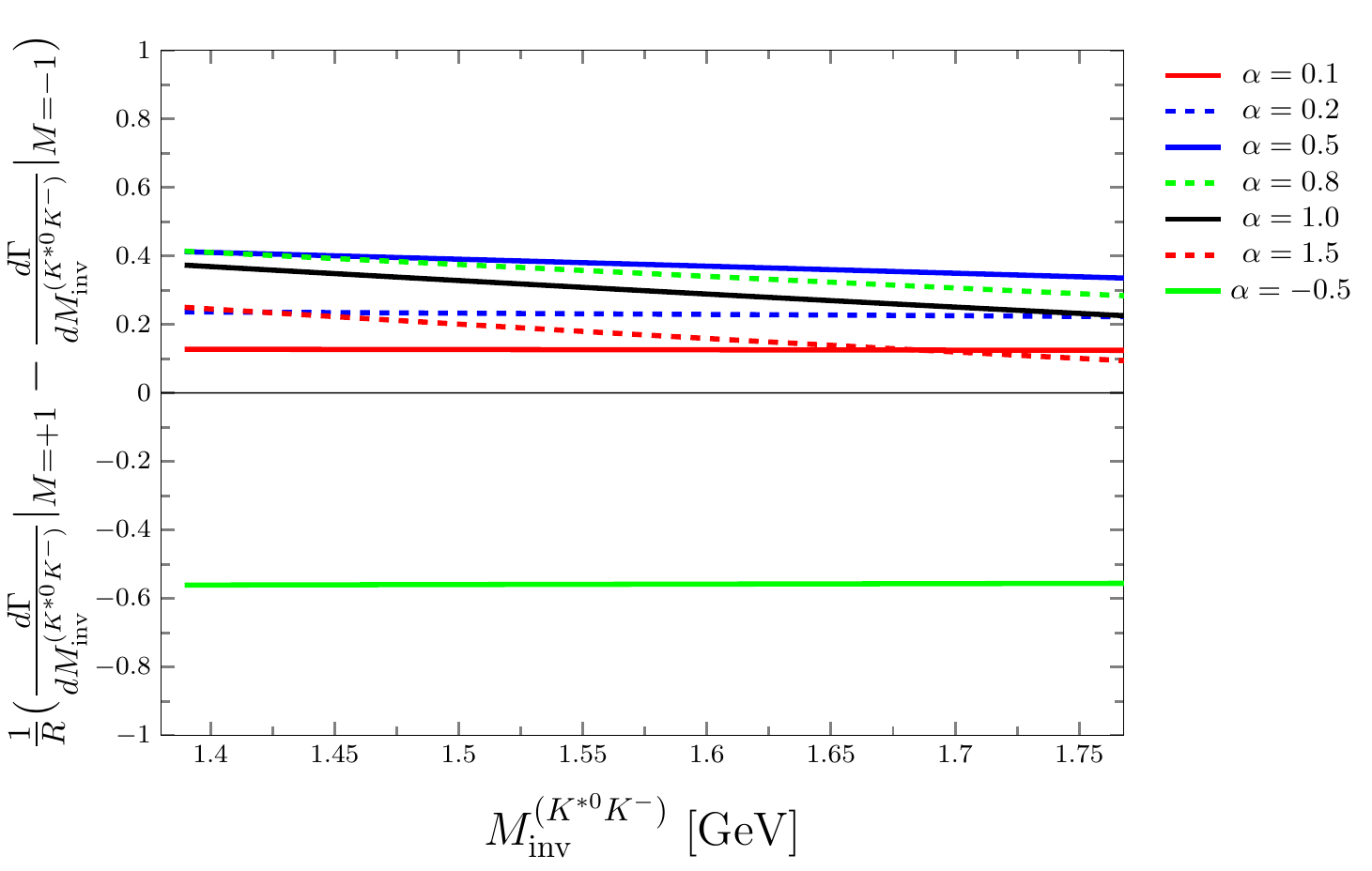}
\caption{ The  difference of  $\frac{d \Gamma}{d M_{\rm inv}^{(K^{*0} K^-)}}|_{M=+1}-\frac{d \Gamma}{d M_{\rm inv}^{(K^{*0} K^-)}}|_{M=-1}$
as a function of $\alpha$, divided by the total differential width $R$.  }
\label{fig:dgalpha}
\end{figure}

 In Fig. \ref{fig:alpha} we show the results for values of $\alpha$ not too different from unity. As we can see, the results of  the polarization amplitudes vary very much
 with $\alpha$. These are large changes that could not scape a precise experimental determination of these magnitudes.  In particular we  find a magnitude  which is extremely sensitive to $\alpha$ which
 is  $\frac{1}{R}\frac{d \Gamma}{d M_{\rm inv}^{(K^{*0} K^-)}}|_{M=+1}-\frac{1}{R}\frac{d \Gamma}{d M_{\rm inv}^{(K^{*0} K^-)}}|_{M=-1}$.
 We show this magnitude in Fig. \ref{fig:dgalpha} for different  values of $\alpha$.
  This magnitude changes even  sign for some value of $\alpha$  and turns out to be most suited to investigate possible deviations of the SM.

\section{Final state interaction and resonance production, connection with other formalisms}

\begin{figure}[ht!]
\includegraphics[scale=0.66]{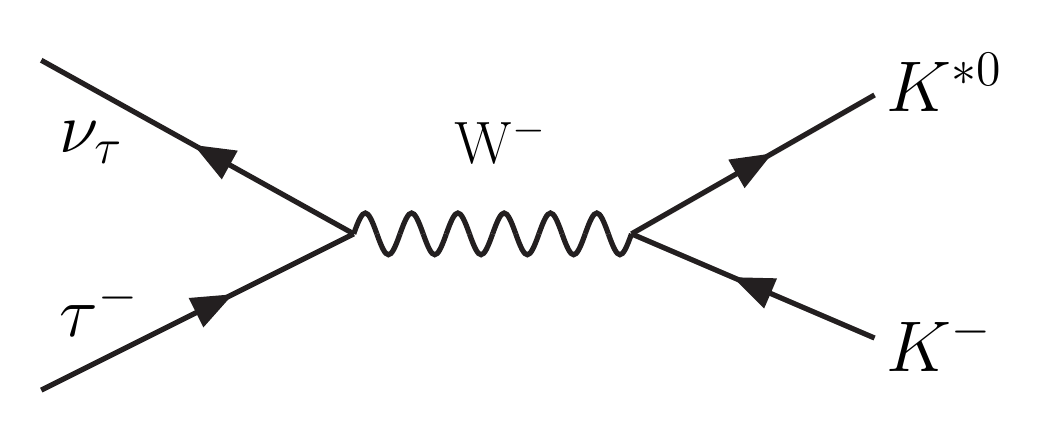}\includegraphics[scale=0.6]{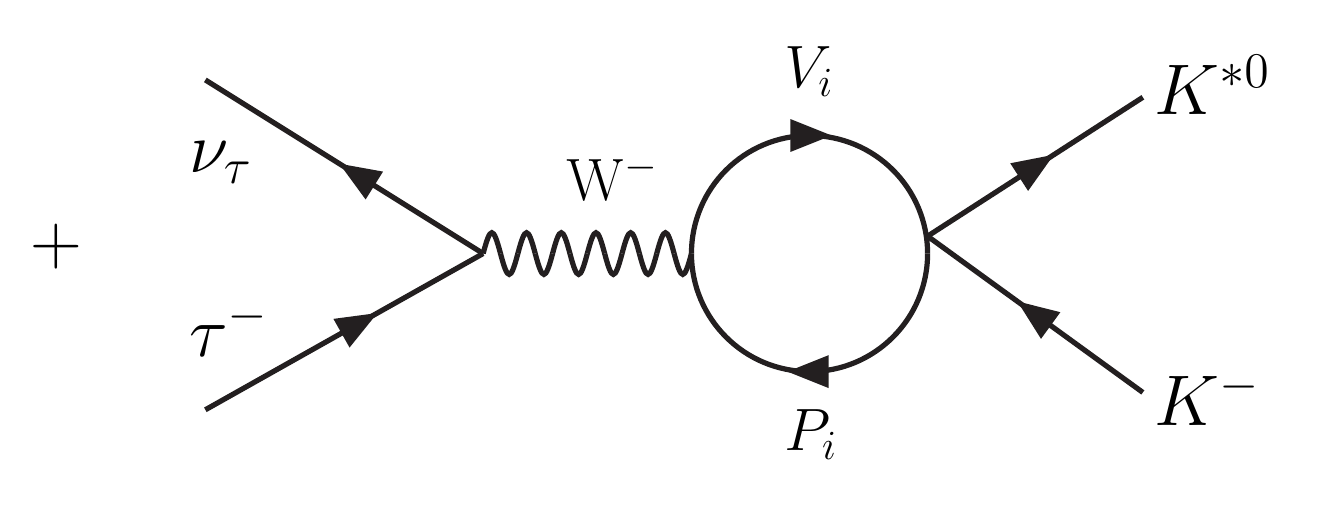}
\caption{Diagrammatic representation of  $K^{*0} K^{-}$ production  with final state interaction of coupled $V_i P_i$ channels.  }
\label{fig:new}
\end{figure}

We have  used a contact interaction for the evaluation  of the $\tau^- \to \nu_{\tau} K^{*0} K^{-}$ amplitudes. However, it is  well known that axial vector resonance, like the
$a_1(1260)$ have contribution  to this process, as evidenced  in the related $\pi^- \rho$ channel \cite{Schael,bali2,Leupold}. From our perspective, the contribution  of these  resonances
appears  considering the final state interaction  of the components that come from the hadronization of $d\bar{u}$. As shown in  \cite{daitau}, the combination stemming from this hadronization
is given by
\begin{eqnarray}\label{eq:VP21}
(V \cdot P)_{21}=\rho^- \left(\frac{\pi^0}{\sqrt{2}}+\frac{\eta}{\sqrt{3}}+ \frac{\eta^{\prime}}{\sqrt{6}} \right)
 + \left( -\frac{\rho^0}{\sqrt{2}}+\frac{\omega}{\sqrt{2}}  \right)\pi^- + K^{*0}K^- \, .
\end{eqnarray}
This means that the  $K^{*0} K^{-}$  in the final state is reached  by the series of diagrams depicted in Fig. \ref{fig:new},
where $V_i P_i$  is any of the coupled channels that appears in  Eq. \eqref{eq:VP21}. This means that we must multiply the elementary amplitude previously evaluated by
\begin{eqnarray}\label{eq:n1}
1 + \sum_i G_i t_{i,K^{*0} K^{-}}  \,
\end{eqnarray}
to obtain the final amplitude, where $G_i$ is the vector-pseudoscalar loop function and $t_{i,K^{*0} K^{-}}$ the transition amplitude from channel $i$ to $K^{*0} K^{-}$. These
magnitudes  are evaluated in \cite{Lutz,Roca,Geng} using the chiral unitary  approach  for the $VP$ interaction in $s$-wave. The important thing for the present work is that
the $t_{ij}$ matrix conserves the spin of the vector, with the vertex ${\bm \epsilon} \cdot {\bm \epsilon'}$, and hence, the polarization  in all the vectors  in Fig. \ref{fig:new}  is the same,
and the interaction factor of Eq. \eqref{eq:n1} is independent of spin. As a consequence,  while all the polarization  amplitudes will  be affected  by this interaction, that generates dynamically
the axial vector resonances, the different polarization  amplitudes are affected by the same factor and in the ratios it cancels exactly.

Next we would like to make connection to the conventional representation of the  amplitudes in terms of invariants that contain the polarization vector of the $K^{*}$. For this purpose we find useful
to use the formalism of  \cite{Caprini,Fajfer} in the related semileptonic decay $ K^* \to l \bar{\nu} K$. By analogy to $B \to D^* \nu l$ in \cite{Caprini,Fajfer} we write
\begin{eqnarray}\label{eq:n2}
&\langle K^*,\lambda, P_{K^*}|{\cal J}_\mu (0)|K, P_K \rangle = \frac{2i V(q^2)}{m_K+m_{K^*}} \epsilon_{\mu\nu\alpha\beta} (\epsilon^{(\lambda)\nu})^* P^{\alpha}_K P^{\beta}_{K^*}
- 2 m_{K^*} A_0(q^2) \frac{\epsilon^{(\lambda)*} \cdot q}{q^2} q_\mu  \nonumber \, \\
& - ({m_K +m_{K^*}})  A_1(q^2) \left[\epsilon^{(\lambda)*}_\mu -\frac{\epsilon^{(\lambda)*} \cdot q}{q^2} q_\mu \right]    \nonumber \, \\
& + A_2(q^2) \frac{\epsilon^{(\lambda)*} \cdot q}{m_K+m_{K^*}}  \left[ (P_K+P_{K^*})_\mu -\frac{m^2_K-m^2_{K^*}}{q^2}q_\mu  \right]  \,,
\end{eqnarray}
where $q=P_{K}-P_{K^*}$ (in the present case since $K^*, K$  are both produced $q=P_{K}+P_{K^*}=P_{\tau}-P_{\nu}$).  In \cite{helicity} the connection of this formalism to our formalism
in terms of the explicit $M=0,\pm 1$ amplitudes is done. In the present case it is unnecessary  to repeat the procedure because we can take advantage of the results found on \cite{helicity}
to justify our results. Indeed, it might appear surprising that the formalism of Eq. \eqref{eq:n2} requires four form factors while we could make  predictions with no free parameters. For this
we can go to  Ref. \cite{helicity} (Fig. 5 and Fig. 10)  where one observes that the  behaviour of the different $M$ contributions to  $\frac{d \Gamma}{d M_{\rm inv}^{(\nu l)}}$ at $M_{\rm inv}^{(\nu l)}/{\rm max}$,
divided by the total differential width $R$ of Eq. \eqref{eq:R},
 is exactly the same in our approach  as for the Standard Model using the empirical values of the four form factors.
 The reason was found in  \cite{helicity} because at $M_{\rm inv}^{(\nu l)}$ maximum  only the   $A_1(q^2)$ amplitude contributes. In addition it was found that this term dominates in a region of
 $M_{\rm inv}^{(\nu l)}$ values close the the maximum (in a range of $500$ MeV where our results are practically indistinguishable from those of the Standard Model). As a consequence of that,  our approach  providing  a slightly  different  functional of  $A_1(q^2)$ than the  Standard Model gave basically  the  same results  of the Standard Model in the ratios of amplitudes, where $A_1(q^2)$  cancels. In the present  case there is not much phase space in the $\tau^- \to \nu_{\tau} K^{*0} K^{-}$  reaction, and the distance of the peak of the invariant  mass distribution  to  $M_{\rm inv}^{(\nu l)}/{\rm max}$ is about
$250$ MeV \cite{bali2}, so it falls well within the range of dominance of $A_1(q^2)$,  which guarantees  that it will  cancel  in ratios, which  can then be obtained  with no parameters as it has
naturally appeared in our approach.

\section{Final remarks}
The $\tau$ decays have been the subject of intense research,  both experimental and theoretical.  Ongoing  discussion continues and there are prospects for construction of new
facilities,  one of them  at Hefei \cite{hefei}, another one at Novosibirsk \cite{novosibirsk}  (see also `` workshop on future tau-charm factory'' \cite{workshop}).
In addition, the new Belle  upgrade will devote  further attentions to the subject \cite{bellebook}. 

Polarization  measurements have been  addressed in related semileptonic  reactions  and can be well adapted to $\tau$  decays. The separation of the polarization amplitudes can be accomplished 
 by looking  at the  decay products of $K^* \to K \pi $ and  angular correlations between the momenta  of these   particles  and the momentum of the $\tau$, $\nu$  
 in the rest frame of $K^* \bar{K}$ \cite{n41,n42,jorge,newbelle}.
 
 Finally  we should also remark  that the form of quark operator  to go beyond  the Standard Model  that we have  adopted  [Eq. \eqref{eq:19}]  does not account for all the large
 variety  of models BSM. It would be closet to the Right-Left handed models used in \cite{nhe,nbaur,German,n36,new1,new2}. However, the fact that we  have seen much sensitivity of the polarization 
results in this subset of BSM models, should be an incentive for the study of these observables  within other models, such that  future experiments can serve to elucidate between them.

\section{Conclusions}
We have performed a calculation of the different  vector meson polarization contribution for a  $\tau^- \to \nu_{\tau} V M $ decay, concretly  $\tau^- \to \nu_{\tau} K^{*0} K^{-}$.
The $M=\pm 1, 0$ third components of the $K^{*0}$ spin are taken  with respect to a frame where the $z$ axis  is chosen in  the neutrino direction. In a first step we evaluate  these contributions
to $\frac{d \Gamma}{d M_{\rm inv}^{(K^{*0} K^{-} )}}$ within the Standard Model and we see that  they differ from each other appreciably.  In view of that we extend the calculations to the
case of one weak hadronic vertex of the type $\gamma^\mu -\alpha\gamma^\mu \gamma_5$, which when $\alpha \neq  1$  can accommodate a sub-set of  models  beyond the Standard Model. We find  that
the results depend strongly  on  $\alpha$ and particularly the magnitude $\frac{d \Gamma}{d M_{\rm inv}^{(K^{*0} K^{-} )}}|_{M=+1} -\frac{d \Gamma}{d M_{\rm inv}^{(K^{*0} K^{-} )}}|_{M=-1} $
divided by the sum of the three contributions ($M=\pm 1, 0$), which changes even sign for some values of $\alpha$.
We also discussed that resonance production in these reactions, mostly through excitation of axial vector resonances,  modifies the contribution of the different $M$ amplitudes but not their ratio.
Furthermore, by making a comparison with the invariant amplitude formalism with explicit polarization vectors, by  analogy  to studies of the semileptonic  decay of  hadrons which involves four
invariant form factors, we found that the present amplitudes are dominated by just the $A_1(q^2)$   form factor, which then cancels in ratios, such that  they can be
calculated  with no free parameters as we have done in our approach. In view of all this, we propose the measurement  of these polarization magnitudes, which are
bound to provide new light on the Standard Model and  possible extensions beyond the Standard Model.

\section*{Acknowledgments}

We wish to express our thanks to Jose Valle, Martin Hirsch, Avelino Vicente, Xiao-Gang He, Juan Nieves and Eliecer Hernandez for useful discussions.
LRD acknowledges the support from the National Natural Science
Foundation of China (Grant No. 11575076) and the State Scholarship Fund of China (No. 201708210057).
This work is partly supported by the Spanish Ministerio
de Economia y Competitividad and European FEDER funds under Contracts No. FIS2017-84038-C2-1-P B
and No. FIS2017-84038-C2-2-P B, and the Generalitat Valenciana in the program Prometeo II-2014/068, and
the project Severo Ochoa of IFIC, SEV-2014-0398 (EO).

\appendix
\section{Evaluation of $\overline{\sum} \sum |t|^2$}
We start from Eqs.~\eqref{eq:L},\eqref{eq:LL} and follow the nomenclature $\bar{L}^{\mu \nu} = \overline{\sum} \sum L^{\mu} L^{\dagger \nu}$ adopted before for simplicity, we have for the leptonic sector
\begin{eqnarray}
\bar{L}^{\mu \nu} \equiv  \overline{\sum} \sum L^{\mu} L^{\dagger \nu} &=& \frac{1}{m_{\tau} m_{\nu}} \left\{ p'^{\mu} p^{\nu} + p'^{\nu} p^{\mu} - g^{\mu \nu} (p' \cdot p)    \nonumber \,\right. \\
 & +&  \left.  i \epsilon^{\alpha \mu \beta \nu} p'_{\alpha} p_{\beta} \right\}.
\end{eqnarray}
Recall that we take the direction of the neutrino as the  quantum $z$ axis and evaluate the different magnitudes in the $K^{*0} K^{-}$  rest  frame.

Thus, for the leptonic plus hadronic matrix elements we have
\begin{equation}
\label{eq:9-1}
\overline{\sum} \sum |t|^2 = \bar{L}^{0 0} M_0 M^*_0 + \bar{L}^{0 i} M_0 N^*_i + \bar{L}^{i 0} N_i M^*_0 +  \bar{L}^{i j} N_i N^*_j.
\end{equation}

\subsection{For $M=0$ }
\begin{enumerate}
\item[1)] $\bar{L}^{00} M_0 M^*_0 $ term
\begin{eqnarray}
\bar{L}^{00} &=& \frac{1}{m_\tau m_\nu}  \left[E_\nu E_\tau + E_\tau E_\nu - (p_\nu \cdot p_\tau) \right]   \nonumber \,\\
&=&\frac{1}{m_\tau m_\nu}  \left( 2 E_\nu E_\tau - E_\tau E_\nu +{\bm{p}_\tau} \cdot {\bm{p}_\nu}  \right)   \nonumber \,\\
&=&\frac{1}{m_\tau m_\nu}  \left( E_\nu E_\tau  +{\bm{p}^2_\nu}  \right)
\end{eqnarray}
Recall that in that frame of reference ${\bm{p}_\tau} = {\bm{p}_\nu} =\bm{p}$.
Using Eq. \eqref{eq:m0}, we obtain
\begin{eqnarray} \label{eq:00M0}
\bar{L}^{00} M_0 M^*_0 = \frac{1}{m_\tau m_\nu} \left(\frac{1}{\sqrt{6}}\frac{1}{4\pi}\right)^2 \left(E_\tau E_\nu +p^2\right) \, ,
\end{eqnarray}

\item[2)]  $\bar{L}^{0i} M_0 N^*_i $ and  $\bar{L}^{i0} N_i M^*_0$ terms
\begin{eqnarray}
\bar{L}^{0i} = \frac{1}{m_\tau m_\nu}  \left[p^0_\nu p^i_\tau + p^0_\tau p^i_\nu - g^{0i}(p^0_\nu \cdot p^i_\tau) +
i \epsilon^{\alpha 0 \beta i} p_{\nu\alpha} p_{\tau\beta}\right] \, ,
\end{eqnarray}
where  $g^{0i}=0$,  and for  $ i \epsilon^{\alpha 0 \beta \nu} $, $\alpha,\beta$ are spatial,  ${\bm{p}_\tau}={\bm{p}_\nu}$, thus
$\epsilon^{\alpha 0 \beta \nu} p_{\nu\alpha} p_{\tau\beta}=0$.

Now let us use the relationship
\begin{eqnarray}
\sum_{i} p^i_{\tau}N^*_i=\sum_{\mu} p_{\tau\mu} N^*_\mu  \, ,
\end{eqnarray}
with $\mu$ in spherical basis, and since we take ${\bm{p}}$ in the $z$ direction,
\begin{eqnarray}
\sum_{i} p^i_{\tau}N^*_i= p_{\tau z} N^*_0  \, ,
\end{eqnarray}
and using Eq.\eqref{eq:nu},
\begin{equation}
N_0=  \frac{1}{\sqrt{3}}\frac{1}{4\pi} {\cal C}(1 1 1; M,0,M) = \frac{1}{\sqrt{3}}\frac{1}{4\pi} {\cal C}(1 1 1; 0,0,0) =0  \, ,
\end{equation}
hence, the $\bar{L}^{0i} M_0 N^*_i $ term gives no contribution, and the same thing happens for  the $\bar{L}^{i0} N_i M^*_0 $ term.
\item[3)] The  $\bar{L}^{ij} N_i N^*_j$ term. We have
\begin{eqnarray}
\bar{L}^{ij} N_i N^*_j=\sum_{\alpha, \beta} (-1)^\alpha \bar{L}^{\alpha \beta}  N_{-\alpha}  N^*_\beta \, .
\end{eqnarray}
with $\alpha, \beta$ in spherical basis.
So for different terms, we obtain
\begin{enumerate}
\item[(a)] since ${\bm{p}_\tau}, {\bm{p}_\nu}$  are in the  $z$ direction, we have
\begin{eqnarray} \label{eq:a}
&(-1)^\alpha (p_{\nu\alpha} p_{\tau\beta}+p_{\nu\beta} p_{\tau\alpha}) N_{-\alpha}  N^*_\beta =(p^2+p^2) N_0 N^*_0   \nonumber \,\\
&= 2 p^2 {\cal C}(1 1 1; 0,0,0){\cal C}(1 1 1; 0,0,0)=0   \,.
\end{eqnarray}
\item[(b)] for the $-g_{ij} (p_\nu \cdot p_\tau) N_i N^*_j$ term, we have
\begin{eqnarray} \label{eq:ijM0}
&-g_{ij} (p_\nu \cdot p_\tau) N_i N^*_j = \delta_{ij} (p_\nu \cdot p_\tau) N_i N^*_j =\sum_{\mu} (p_\nu \cdot p_\tau) N_\mu N^*_\mu \nonumber \,\\
&=\sum_{\mu}\frac{1}{3} \left(\frac{1}{4\pi}\right)^2 {\cal C}^2(1 1 1; 0,-\mu,-\mu) (p_\nu \cdot p_\tau) \nonumber \,\\
&=\frac{1}{3} \left(\frac{1}{4\pi}\right)^2 (\frac{1}{2}+0+\frac{1}{2}) (p_\nu \cdot p_\tau)=\frac{1}{3} \left(\frac{1}{4\pi}\right)^2 (p_\nu \cdot p_\tau) \,.
\end{eqnarray}
\item[(c)] for the $ \epsilon^{\alpha i \beta j} p_{\nu\alpha} p_{\tau\beta} N_i N^*_j$ term, since $\alpha$ or $\beta$ must be zero, so we take $\alpha=0$, then this term becomes
$ E_\nu \epsilon^{i \beta j} p_{\tau\beta} N_i N^*_j$, after taking $\beta=3$ which means  $z$ direction or third component, and using Eqs. \eqref{eq:m0} and \eqref{eq:nu}, we obtain
\begin{eqnarray} \label{eq:b}
 & i E_\nu \epsilon^{i3j} p N_i N^*_j = - i E_\nu \epsilon^{ 3 i j} p N_i N^*_j = -i E_\nu p(N_1 N_2^*-N_2 N_1^*)=-i i E_\nu p (N_{-}N_{-} -N_{+} N_{+})  \nonumber \,\\
 &= E_\nu p \left[ {\cal C}^2(1 1 1; 0,+1,+1)-{\cal C}^2(1 1 1; 0,-1,-1) \right]=0   \, ,  ~~~~
\end{eqnarray}
where we use the expressions of Eq.\eqref{eq:spbs} for  $N_{+}$ and $N_{-}$ .
\end{enumerate}
\end{enumerate}
Altogether, from Eqs. \eqref{eq:00M0} and \eqref{eq:ijM0}, we obtain
\begin{eqnarray}
\overline{\sum} \sum |t|^2 &=& \frac{1}{m_\tau m_\nu} \left[\frac{1}{6}\left(\frac{1}{4\pi}\right)^2 \left(E_\tau E_\nu +{\bm{p}^2_\nu} \right)+\frac{1}{3} \left(\frac{1}{4\pi}\right)^2 (p_\nu \cdot p_\tau) \right] \nonumber \,\\
&=&\frac{1}{m_\tau m_\nu} \frac{1}{6}\left(\frac{1}{4\pi}\right)^2 \left[\left(E_\tau E_\nu +p^2 \right)+ 2 \left(E_\tau E_\nu -p^2 \right) \right]
\end{eqnarray}

\subsection{for $M=\pm 1$ }

\begin{enumerate}
\item[1)] $\bar{L}^{00} M_0 M^*_0 $  terms

The same result as above \eqref{eq:00M0},  because $M_0$ is the same for all $M$,
\begin{eqnarray}\label{eq:x1}
\bar{L}^{00} M_0 M^*_0 = \frac{1}{m_\tau m_\nu} \left(\frac{1}{\sqrt{6}}\frac{1}{4\pi}\right)^2 \left(E_\tau E_\nu +p^2 \right) \, ,
\end{eqnarray}
\item[2)] $\bar{L}^{0i} M_0 N^*_i $ and $\bar{L}^{i0} N_i M^*_0$   terms
\begin{eqnarray} \label{m1tem}
\bar{L}^{0i} M_0 N^*_i=\frac{1}{m_\tau m_\nu}  \left(p^0_\nu p^i_\tau + p^0_\tau p^i_\nu \right) M_0 N^*_i = \frac{1}{m_\tau m_\nu} \left\{(E_\tau +E_\nu) \sum_{\mu} p_\mu N^*_\mu\right\} M_0 \, ,~~~~~
\end{eqnarray}
By taking the third  $p_3$ component ($z$ direction in spherical basis), Eq. \eqref{m1tem} becomes
\begin{eqnarray}\label{m1tem2}
\bar{L}^{0i} M_0 N^*_i &=& \frac{1}{m_\tau m_\nu} \left\{(E_\tau +E_\nu) \,p\, \frac{1}{\sqrt{6}} \frac{1}{4 \pi} \frac{1}{\sqrt{3}} \frac{1}{4 \pi} (-1)^0 {\cal C}(1 1 1; M,0,M) \right\}   \nonumber \,\\
&=& \frac{1}{m_\tau m_\nu} \left\{(E_\tau +E_\nu)  \frac{1}{6} \left(\frac{1}{4\pi}\right)^2  \,p\ M \right\} \, ,~~~~~
\end{eqnarray}
and $\bar{L}^{i0} N_i M^*_0$ is the complex conjugate of the former one of  Eq. \eqref{m1tem2}.

Altogether,  we obtain
\begin{eqnarray} \label{i0m1}
\bar{L}^{0i} M_0 N^*_i+\bar{L}^{i0} N_i M^*_0 = \frac{1}{m_\tau m_\nu} \frac{1}{3} \left(\frac{1}{4\pi}\right)^2 (E_\tau +E_\nu) \,p\ M \, .
\end{eqnarray}
\item[3)] The  $\bar{L}^{ij} N_i N^*_j$ term
\begin{enumerate}
\item[(a)] since  ${\bm{p}_\tau}, {\bm{p}_\nu}$  are in the $z$ direction, in analogy to  Eq. \eqref{eq:a},  but having here  $M=\pm 1$, we  have
\begin{eqnarray}\label{eq:x2}
&\frac{1}{m_\tau m_\nu} (-1)^\alpha (p_{\nu\alpha} p_{\tau\beta}+p_{\nu\beta} p_{\tau\alpha}) N_{-\alpha}  N^*_\beta   \nonumber \,\\
& = \frac{1}{m_\tau m_\nu} (p^2+p^2) N_0 N^*_0  \nonumber \,\\
&= \frac{1}{m_\tau m_\nu} 2 p^2  \frac{1}{3} \left(\frac{1}{4\pi}\right)^2   {\cal C}^2(1 1 1; M,0,M) =\frac{1}{m_\tau m_\nu} \frac{1}{3} \left(\frac{1}{4\pi}\right)^2   p^2  \,.
\end{eqnarray}
\item[(b)] for $-g_{ij} (p_\nu \cdot p_\tau) N_i N^*_j$ term,

The same as Eq.\eqref{eq:ijM0} but here for  $M=\pm 1$, we  have
\begin{eqnarray}\label{eq:x3}
&-\frac{1}{m_\tau m_\nu} g_{ij} (p_\nu \cdot p_\tau) N_i N^*_j = \frac{1}{m_\tau m_\nu} \delta_{ij} (p_\nu \cdot p_\tau) N_i N^*_j = \frac{1}{m_\tau m_\nu} \sum_{\mu} (p_\nu \cdot p_\tau) N_\mu N^*_\mu \nonumber \,\\
&= \frac{1}{m_\tau m_\nu} (p_\nu \cdot p_\tau)  \frac{1}{3} \left(\frac{1}{4\pi}\right)^2 \sum_{\mu}  {\cal C}^2(1 1 1; M,-\mu,M-\mu)  \nonumber \,\\
&=\frac{1}{m_\tau m_\nu} (p_\nu \cdot p_\tau)  \frac{1}{3} \left(\frac{1}{4\pi}\right)^2  \,. ~~~~~
\end{eqnarray}
\item[(c)] for the $ i \epsilon^{\alpha i \beta j} p_{\nu\alpha} p_{\tau\beta} N_i N^*_j$ term, $\alpha$ or $\beta$ must be zero.  \\
First,  if  we take $\alpha=0$,
\begin{eqnarray}\label{alpha}
& i \epsilon^{\alpha i \beta j} p_{\nu\alpha} p_{\tau\beta} N_i N^*_j = i E_\nu  \epsilon^{ i \beta j}  p_{\tau\beta} N_i N^*_j\nonumber \,\\
&= - i E_\nu  \epsilon^{ 3 i j}  p N_i N^*_j  = i i  E_\nu p (N_{-}N_{-} -N_{+} N_{+}) \, ,
\end{eqnarray}
then, Eq.\eqref{alpha}  becomes
\begin{eqnarray} \label{eq:aa}
 - E_\nu p  \frac{1}{3} \left(\frac{1}{4\pi}\right)^2  \left[{\cal C}^2(1 1 1; M,+1,M+1)-{\cal C}^2(1 1 1; M,-1,M-1)\right] \,.
\end{eqnarray}
Second,  if we take $\beta=0$,
\begin{eqnarray}\label{beta}
& i \epsilon^{\alpha i 0 j} p_{\nu\alpha} p_{\tau\beta} N_i N^*_j = i E_\tau  \epsilon^{\alpha i j}  p_{\nu\alpha} N_i N^*_j\nonumber \,\\
&=  i E_\tau  \epsilon^{ 3 i j}  p N_i N^*_j  = E_\tau p (N_{-}N_{-} -N_{+} N_{+}) \, ,
\end{eqnarray}
then  Eq.\eqref{beta}  becomes
\begin{eqnarray}\label{eq:bb}
 E_\tau \,p \, \frac{1}{3} \left(\frac{1}{4\pi}\right)^2  \left[{\cal C}^2(1 1 1; M,+1,M+1)-{\cal C}^2(1 1 1; M,-1,M-1)\right] \,.
\end{eqnarray}
\end{enumerate}
then adding  the results of Eq. \eqref{eq:aa} and \eqref{eq:bb}, calculating the CGC explicitly we find including the factor $\frac{1}{m_\tau m_\nu}$
\begin{eqnarray}  \label{eq:x4}
\frac{1}{m_\tau m_\nu} i \epsilon^{\alpha i \beta j} p_{\nu\alpha} p_{\tau\beta} N_i N^*_j =\frac{1}{m_\tau m_\nu} \frac{1}{6} \left(\frac{1}{4\pi}\right)^2 (E_\nu -E_\tau) \,p\ M \, .
\end{eqnarray}
Altogether, from  Eqs. \eqref{eq:x1}, \eqref{i0m1}, \eqref{eq:x2},  \eqref{eq:x3}, \eqref{eq:x4}, we obtain
\begin{eqnarray}
\overline{\sum} \sum |t|^2 =\frac{1}{m_\tau m_\nu} \frac{1}{6}\left(\frac{1}{4\pi}\right)^2 \left\{3 E_\tau E_\nu + p^2+  \left(3 E_\nu + E_\tau \right) p M \right\} \, .
\end{eqnarray}
\end{enumerate}

For the case of $\alpha \neq 1 $, all we must do is to consider that now $N_i$   gets multiplied by $\alpha$ and we readily get the  new equations.

\end{document}